\documentclass[twocolumn,showpacs,preprintnumbers,amsmath,amssymb]{revtex4}
\usepackage[amssymb]{SIunits}
\usepackage{amsmath}
\usepackage{graphicx}
\usepackage{color}
\usepackage{bm}	
\usepackage{dcolumn}
\usepackage{amssymb}

\begin{document}
\pacs{XXX}

\title{Spatial Mapping and Manipulation of two tunnel coupled Quantum Dots}

\author{Magdalena Huefner\footnote{Present address: Department of Physics, Harvard University, Cambridge, Massachusetts 02138, USA}}
\email{huefner@physics.harvard.edu}
\affiliation{
Solid State Physics Laboratory, ETH Z\"urich, 8093 Z\"urich,
Switzerland}

\author{Bruno Kueng}
\affiliation{
Solid State Physics Laboratory, ETH Z\"urich, 8093 Z\"urich,
Switzerland}

\author{Stephan Schnez}
\affiliation{
Solid State Physics Laboratory, ETH Z\"urich, 8093 Z\"urich,
Switzerland}

\author{Klaus Ensslin}
\affiliation{
Solid State Physics Laboratory, ETH Z\"urich, 8093 Z\"urich,
Switzerland}

\author{Thomas Ihn}
\affiliation{
Solid State Physics Laboratory, ETH Z\"urich, 8093 Z\"urich,
Switzerland}

\author{Matthias Reinwald}
\affiliation{
Institut f\"ur Experimentelle und Angewandte Physik
Universitaet Regensburg, 93040 Regensburg, Germany}

\author{Werner Wegscheider}
\affiliation{
Solid State Physics Laboratory, ETH Z\"urich, 8093 Z\"urich,
Switzerland }

\date{\today}

\begin{abstract}
 The metallic tip of a scanning force microscope operated at $300\text{ }\milli\kelvin$ is used to locally induce a potential in an fully controllable double quantum dot defined via local anodic oxidation in a GaAs/AlGaAs heterostructure. Using scanning gate techniques we record spatial images of the current through the sample for different numbers of electrons on the quantum dots, i.e. for different quantum states. Owing to the spatial resolution of current maps, we are able to determine the spatial position of the individual quantum dots, and investigate their apparent relative shifts due to the voltage applied to a single gate.
\end{abstract}
\maketitle

\section{Introduction}\label{sec_intro}

Most quantities measured in pure transport experiments, such as the current through a nanostructure, remain macroscopic and contain little spatial information. To justify assumptions about the potentially spatially non-uniform current flow in nanostructures it is highly useful to employ a probing technique which offers spatial resolution. Scanning gate microscopy, where a metallic tip is used to capacitively couple to the nanostructure acting as a freely movable gate, is a powerful local probing technique. It has been applied to a broad range of materials from 2DEG-based semiconductors \cite{Topinka_Nature_2001,Pioda_PRL_2004,Steele_PRL_2005,Crook_Science_2006,Hackens_NaturePhys_2006} over InAs nanowires \cite{Bleszynski_PRB_2008} and graphene \cite{Bleszynski_condmatt_2009,Schnez_Scheissteil_2010} to superconductors \cite{huefner_PRB_2009}. While it has been applied to investigate accidentally formed multiple dots \cite{Bleszynski_NanoLett_2007,Woodside_science_2002} no investigations of controllable, intentionally formed double dots have been reported to date.  \par

Here we show that we can control single electrons on a double quantum dot by moving a metallic tip across the structure. Spatial current maps allow us to gain insight into the topographic position of the charge distribution of the two quantum dots themselves.\par

\section{Sample and Setup}\label{sec_SampleNSetup}
\begin{figure}
  \includegraphics[width=8.25cm]{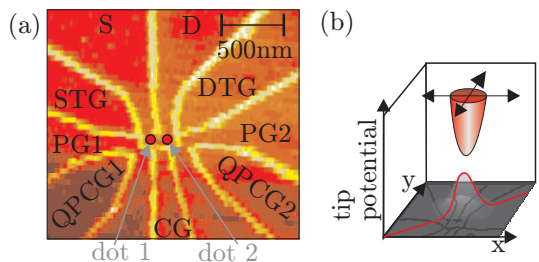}
  \caption{(a) Topography of the investigated structure. The SFM image is recorded at a temperature of $1.7 \text{ }\kelvin$ before carrying out the presented measurements. The insulating oxide lines appear as bright yellow protrusions. The regions below which the conducting 2DEG is located are shown in darker red. The positions where the individual quantum dots are expected to form due to the confinement of the electrons by the oxide lines are marked with red circles. (b) Schematic illustration of the local tip induced potential (grey). The tip is moved in x- and y-direction at constant height above the sample surface.}
  \label{fig_SampleNSetup}
\end{figure}

The double quantum dot under investigation has been fabricated on an AlGaAs-GaAs heterostructure, which contains a two-dimensional electron gas (2DEG) $34 \text{ }\nano\meter$ below the surface. The 2DEG has a density of $n_s=5\times 10^{15}\text{ }\meter^{-2}$ and a mobility of $\mu=40\text{ }\meter^2/\volt\second$ at $4.2\text{ }\kelvin$. The double quantum dot structure has been predefined by local anodic oxidation \cite{Fuhrer_SuperMicro2002} at room temperature. With this technique oxide lines are formed on the surface, below which the 2DEG is locally depleted. Thus the oxide lines electrically separate adjacent 2DEG areas. Figure \ref{fig_SampleNSetup}(a) shows a scanning force microscope (SFM) scan of the investigated structure recorded at a temperature of $1.7\text{ }\kelvin$. The oxide lines are seen as bright protrusions. The positions at which the quantum dots are expected to form are marked with two red dots labeled "dot 1" and "dot 2". The double quantum dot is electrically contacted via tunneling barriers to source (S) and drain (D). The tunneling barriers between source and dot 1 as well as between dot 2 and drain can be tuned with the gates marked STG and DTG respectively. The levels inside each dot can be shifted by applying a voltage to the 2DEG-areas marked PG1, PG2, QPCG1 and QPCG2. The central gate (CG) tunes the interdot coupling.\par
\par
All experiments shown here are performed with a SFM in a $^3$He cryostat with a an electronic temperature of about $650\text{ }\milli\kelvin$. The scanning force sensor consists of an electrochemically etched PtIr tip with an initial radius of about $50\text{ }\nano\meter$, mounted on a piezoelectric tuning fork. The measurements presented in this work are recorded with a clean single tip. Figure \ref{fig_SampleNSetup}(b) schematically shows the tip-induced local electrostatic potential (tip potential) in the sample that can be moved freely in x-y-direction. The coordinates x and y indicate the lateral tip position. The tip-surface distance $d$ for all current maps presented here is $10\text{ }\nano\meter$ unless stated otherwise. The tip induced potential is about $15\text{ }\milli\electronvolt$ for $d=60\text{ }\nano\meter$ . Investigations of the interaction potential between the tip and the sample in previous measurements \cite{Pioda_PRL_2004,huefner_PRB_2009} have shown that even at zero volt tip voltage there is still a noticeable potential induced in the sample. Moving the tip potential in real space across the double quantum dot shifts the electrochemical potential of quantum states. We record so-called current maps, which show the current through the double dot as a function of the tip position $(x,y)$. No current flows between the tip and to the 2DEG.\par

\section{Charge stability diagram in real space}\label{sec_BasicSG}
\begin{figure}
  \includegraphics[width=8.25cm]{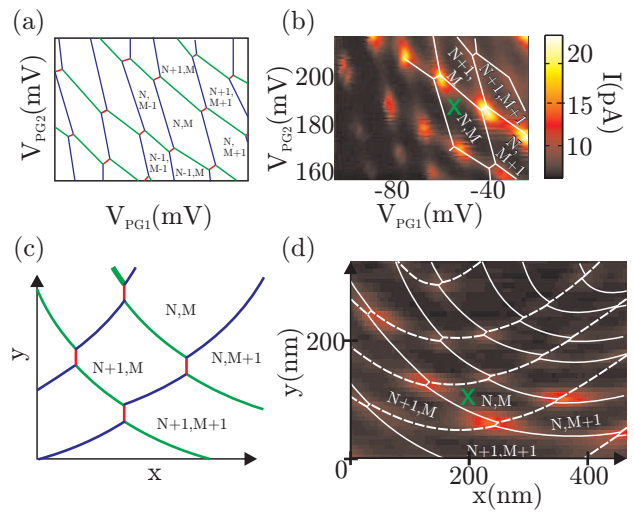}\\
  \caption{(a) Schematic of the hexagon pattern formed when
  measuring $I(V_{PG1},V_{PG2})$ in the double dot regime. The
  letters $i$ and $j$ represent the number of electrons in
  dot 1 and dot 2 respectively. (b) shows the charge stability diagram for a source-drain voltage of
  $210\text{ }\micro\volt$.
  Panel (c) schematically shows the expected scanning gate pattern. The hexagons displayed in (a) are distorted in real space.
  (d) Current map of the double dot. The tip is scanned
  at a tip-sample distance $d=10\text{ }\nano\meter$ above the surface. Pairs of triple points are visible as regions of enhanced current. White lines are used as visual aids to highlight the underlying deformed hexagonal pattern. Labels $(i,j)$ mark the same electron numbers on the dot as in the charge stability diagram shown in (b).
  The green cross marks the gate
  settings used for the current map shown in (d). The green cross in (d) marks
  the position at which the tip was located to record the measurements shown in (b). The green cross in (b) marks the voltage settings used to record the current map in (d).
  Measurement settings: $V_{CG}=120\text{ }\milli\volt$, $V_{STG}=V_{DTG}=-40\text{ }\milli\volt$.}
  \label{fig_SGDD_outer}
\end{figure}

The charge stability diagram of a double quantum dot displays hexagonal patterns as schematically shown in Fig. \ref{fig_SGDD_outer}(a). This pattern mainly consists of lines with two different slopes marked in blue and green. Along the green lines, a quantized energy level of dot 1 is in resonance with source and drain electrochemical potentials. Analogously along the blue lines an energy level of dot 2 is in resonance with the electrochemical potential of source and drain. We can attribute a certain number of electrons on each dot to given gate voltage regions as marked by letters $(i,j)$ in the schematics and  in the corresponding measurement in Fig. \ref{fig_SGDD_outer}(a,b).\par

Panel (b) shows the current through the double quantum dot as a function of the voltage $V_{PG1}$ applied to PG1 and $V_{PG2}$ applied to PG2. The honeycomb pattern characteristic for the charge stability diagram of a double dot can be observed \cite{Wiel_RevModPhys_2002,Reed_Book_1989,Hanson_PvModPhys_2007}. Since the electronic temperature is larger than the mutual charging energy, pairs of triple points form a single current peak. Some pairs of triple points are connected by cotunneling lines of nonzero current. Along such a line, a level of only one quantum dot is aligned with the Fermi energy of source and drain, and cotunneling through the other dot occurs. Both quantum dots have a charging energy of $E_C=3\text{ }\milli \electronvolt$ and the mutual charging energy is $E_{CM}=0.4\text{ }\milli \electronvolt$ and an approximate electron number of 80 electrons. These energy values have been determined from charge stability diagrams taken during a different cooldown cycle at a temperature of $40\text{ }\milli\kelvin$.\par

\begin{figure}
  \includegraphics[width=8.25cm]{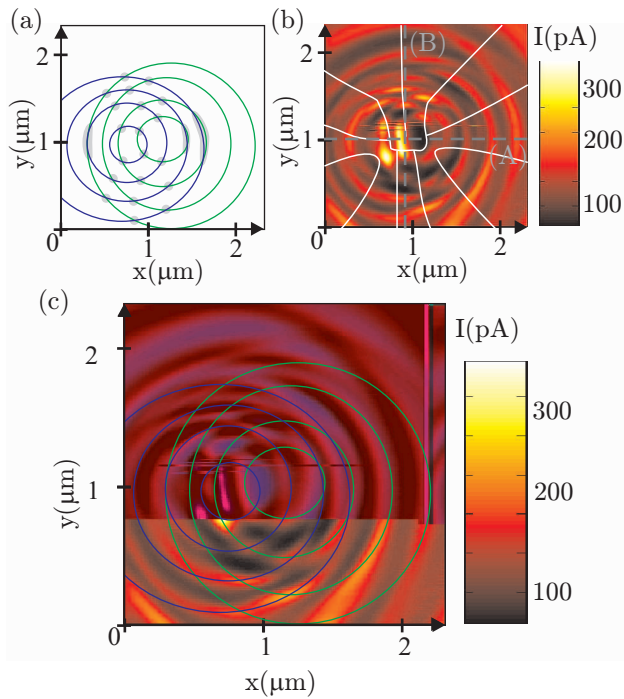}\\
  \caption{Panel (a) schematically displays the two sets of Coulomb rings, one set for each dot. The crossings of the two sets of rings are highlighted with grey dots. At those positions the chemical potential of both dots is in resonance with the leads' chemical potentials. At the corresponding tip positions current can flow through the structure. Panel (b) shows a current map of the double dot.
  The tip is scanned $d=60\text{ }\nano\meter$ above the surface. The oxide lines
  that define the structure are schematically shown as white lines.
  Panel (c) shows a superposition of the images shown in panel (a) and (b).
  }
  \label{fig_SGDD_inner}
\end{figure}

When the current map of a single quantum dot is recorded, concentric rings of conductance resonances are observed \cite{Kicin_NewJPhys2005,Pioda_PRL_2004,Fallahi_NanoLett_2005,huefner_PRB_2009}. Each ring of enhanced current marks tip positions for which the leads chemical potential and one of the dots energy levels are equal. Schematically this corresponds to one set of ellipses in real space, like the green set shown in Fig. \ref{fig_SGDD_inner}(a). When measuring a double dot, each dot will provide its own set of ring-shaped conductance resonances as shown by the blue and green sets. However, current will only flow when both dots are in resonance, i.e. when both sets of rings intersect. These regions which form a distorted hexagon pattern are highlighted in grey. \par

Figure \ref{fig_SGDD_inner}(b) shows the current map of the double dot for the case where both individual dots are equally pronounced. We observe a distorted hexagon pattern corresponding to that schematically shown in Fig. \ref{fig_SGDD_inner}(a): Two sets of rings that are superimposed and lead to a finite current where they intersect. Taking a closer look, each grey spot in Fig. \ref{fig_SGDD_inner} in principle splits into a pair of triple points. This splitting is not observed in the experiment because the mutual charging energy is smaller than the thermal energy at these electronic temperatures. \par

The centers of the two sets of rings are $250\text{ }\nano\meter$ apart, which corresponds to the separation with which the dots are expected to form from the geometric confinement by the oxide lines. On average we observe a tendency that the current decreases, when the tip is moved closer to the double dot, indicating that the tip-potential is repulsive for electrons. At sufficiently negative gate voltage settings, we are even able to completely suppress the conductance of the the double dot with the tip.\par

Figures \ref{fig_SGDD_outer}(c,d) show a more detailed current map taken at a lateral distance of about $y=-1\text{ }\micro\meter$  from the center of the individual dots, where the distorted hexagon pattern is clearly visible. Along each ring section, distinct spots of enhanced current, corresponding to the pairs of triple points, can be distinguished. The hexagonal pattern is highlighted. It is significantly distorted compared to the corresponding measurement in the plane of two gate voltages because the tip-dot lever arms depend strongly on the tip position. For four hexagon regions, the number $(i,j)$ of electrons for each dot is indicated. The green cross in the image indicates the position at which the tip was located to record the charge stability diagram shown in Fig. \ref{fig_SGDD_outer}(b), whereas the green cross in Fig. \ref{fig_SGDD_outer}(b) marks the voltage settings used to record this current map. Therefore the labeled hexagons in Fig. \ref{fig_SGDD_outer} (b) and (d) correspond to the same charge states of the double quantum dot system.\par

To summarize, our observations are the following: First, we are able to distinguish two sets of rings in the current maps, both of which are concentric around one of the two quantum dots. We are therefore able to directly determine the apparent electronic separation of the two dots in the sample plane. Second, by combining charge stability diagrams with conductance maps, we are able to assign a certain electron number $(i,j)$ on the double dot to a spatial position of the tip. Finally, we observe a high number of Coulomb peaks (more than 30  in a $1.9\text{ }\micro\meter\times 1.9\text{ }\micro\meter$ scan frame depending on, among other things, the tip-sample separation $d$), letting us conclude that the quality of our tip is excellent.
\section{Transition from Single to Double dot}\label{sec_SDvsDD}
In addition to tuning the double dot with the plunger gates PG1 and PG2, we can also tune the tunnel barriers using STG and DTG. All previous measurements are performed in a region where both barriers are tuned to the tunneling regime, leading to a display of Coulomb blockade in both dots. Using one of the tunnel barrier gates, we can tune only one dot to the Coulomb blockade regime, whereas the other dot is open, forming an extension of the respective contact. With this method we can confine either dot 1, or dot 2 or both dots simultaneously in this double-dot geometry.\par

\begin{figure}
  \includegraphics[width=8.25cm]{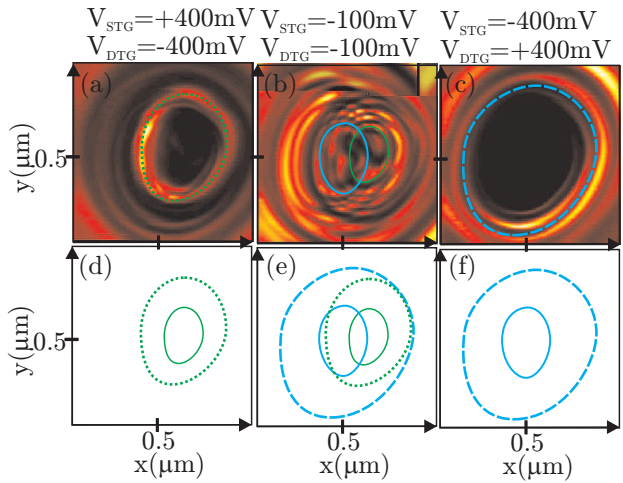}\\
  \caption{Current maps for different tunnel gate voltages
  $V_{STG}$ and $V_{DTG}$. By changing the voltages applied to the
  gates controlling the tunnel barriers we can open the tunnel
  barriers, to lessen confinement of one of the dots. Panel (a) shows the
  case where the tunnel barrier close to the gate DTG is opened,
  leaving only dot 2 as a single dot. Panel (c) shows the
  opposite case, where only dot 1 is formed. When both
  tunnel barriers are reasonably closed, a double dot is formed as
  seen in (b). A single conductance ring is marked with a green
  dotted line in (a). Analogously, one single resonance is traced as a
  blue dashed line in (c). For the case of the double dot in (b),
  one resonance is traced for each dot with a solid line. The second row, (d) through
  (f), shows only these traced resonances. We can see that when the
  tunnel barrier DTG is opened, we observe a single dot at the same
  position as dot 1 of the double dot system in (b). When the tunnel
  barrier STG is opened, we observe only a single dot at the position
  of dot 2 the double dots seen in (f).
  Measurement settings: $V_{PG2}=190\text{
  }\milli\volt$, $V_{PG1}=100\text{ }\milli\volt$, $V_{SD}=200\text{
  }\micro\volt$.
  }
  \label{fig_SG_DD_to_SD}
\end{figure}

Figure \ref{fig_SG_DD_to_SD} (a) through (c) show current maps recorded for these three regimes. For (a), only dot 2 is confined by the central tunneling barrier and the tunneling barrier near DTG. We observe only a single set of concentric conductance rings. One of these rings is highlighted with a green dotted line and also shown in (d). In Fig. \ref{fig_SG_DD_to_SD}(b), we observe clear double-dot behavior. We recognize two sets of concentric rings. Two single resonances are highlighted for clarity. The solid, blue line marks a resonance of dot 1 and the solid, green line marks a resonance of dot 2. When only the dot 1 is confined we observe again single conductance rings as visible in (c). The single resonance traced with a blue, dashed line is shown in Fig. \ref{fig_SG_DD_to_SD} (f). Comparing the 4 traced resonances from Fig. \ref{fig_SG_DD_to_SD} (a-c) in Fig. \ref{fig_SG_DD_to_SD} (e), we observe that the centers of the two blue resonances coincide as do the centers of the two green resonances. These measurements demonstrate that we can establish the position of each single dot in real space. Furthermore, we learn that the position of such a single dot is not changed above the spatial resolution of this measurement, when a second dot is established close to it.

\section{Influence of the gates in real space on the single dots}\label{sec_gates}

In order to investigate the influence of the plunger gate voltages on the double dot, we will now show single line scans, where each line scan is recorded at a different gate voltage. This leads to measurements of the current $I$ versus a spatial direction such as x or y and one gate voltage ($V_{PG1}$, $V_{PG2}$, $V_{CG}$, $V_{STG}$, $V_{DTG}$). These measurements contain the information of a series of current maps in a very condensed way. The real space axes are chosen as follows: One line labeled (A) in Fig. \ref{fig_SGDD_inner}(b) traverses dot 1 and dot 2 in x-direction. The other line labeled (B) in Fig. \ref{fig_SGDD_inner}(b) is the symmetry axis of the structure in y-direction. \par

\begin{figure*}
  \includegraphics[width=17.78cm]{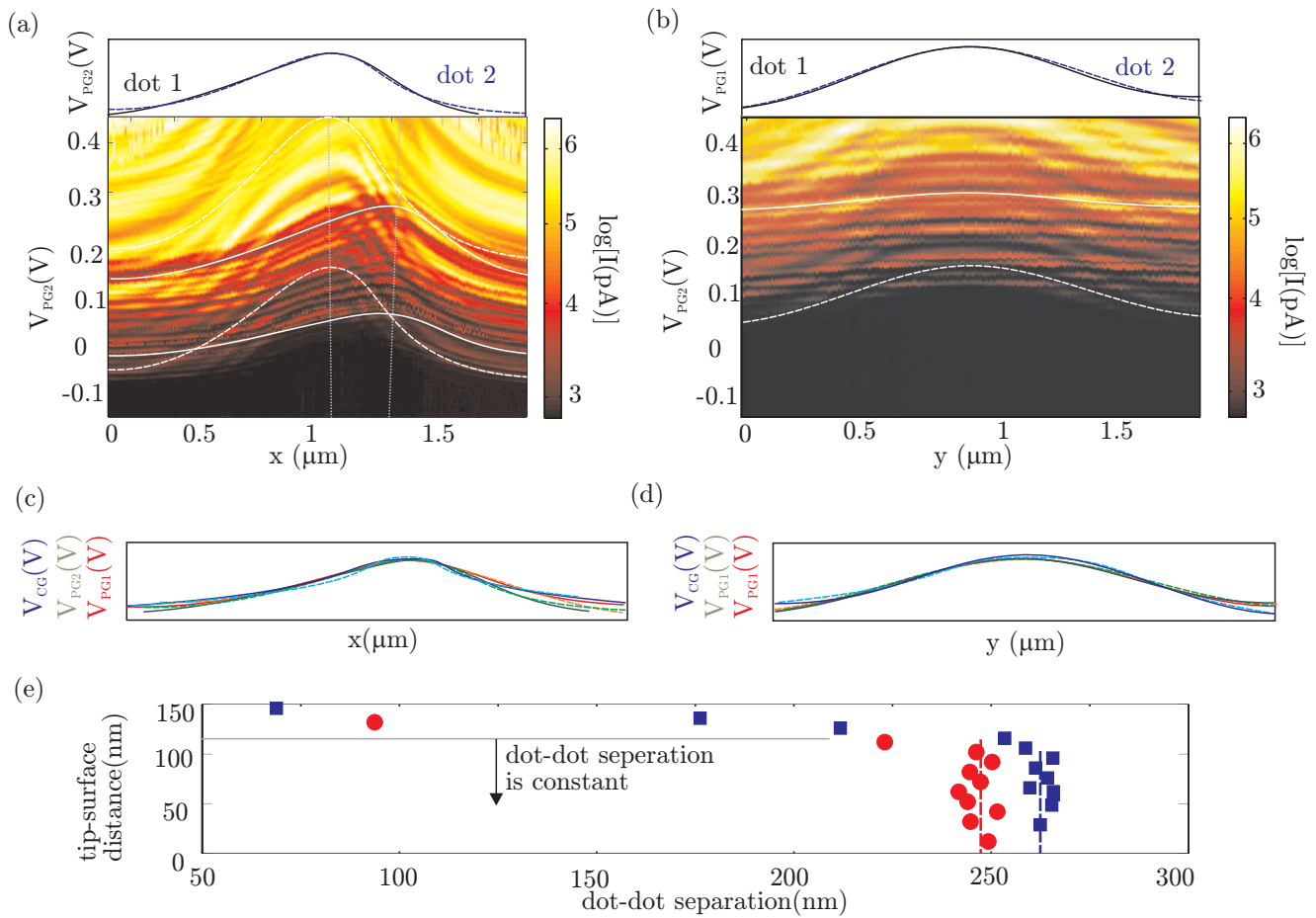}\\
  \caption{Current through the double quantum dot when the tip is
  scanned across a line above the sample. The lines for (a) and (b) are marked with (A) and (B) in Fig.
  \ref{fig_SGDD_inner}. At each point along a
  line the voltage $V_{PG2}$ was swept. The white
  lines trace one particular resonance for each dot.
  The headers are explained in the text. The grey dashed lines in (a) show the shift of the resonance in space due to the changed gate voltage.
  (c) and (d) show an overlay of the profiles obtained from the measurements for different gates.
  (e) the dot-dot separation for different tip-surface separations as derived from a series of scanning gate measurements. We show two tip-surface series of current maps. One is marked with red circles, the second one taken at more positive gate voltages is marked with blue squares.}
  \label{fig_Grid_allGates}
\end{figure*}

Figure \ref{fig_Grid_allGates} (a) shows $I(V_{PG2}, x)$. We observe resonances in the current, where we call a single resonance a profile. One profile presents the line of a constant chemical potential in one dot, while in resonance with the leads. Two sets of profiles are visible: one for dot 1 and a second for dot 2. Two profiles of the right set are traced with solid white lines. Two profiles of the left set are traced with white dashed lines. The left profiles belong to dot 1, whereas the right profiles can be attributed to dot 2. Shifting the profiles along the lateral axis and scaling each of these profiles collapses them onto the same curve (see top panel). This confirms that the two dots sense the same tip-induced potential. When increasing the gate voltage, we observe that the amplitude of the profiles increases by a factor of 3 over the measured voltage range. This is due to the gate voltage dependence of the lever arm. For higher gate voltages the lever arm becomes smaller, therefore the voltage increase needed to  cause a particular energy shift of a quantum dot level becomes larger, leading to profiles with a bigger amplitude. Furthermore we observe that the amplitude of the profile of dot 2 is a factor of two to three smaller, than the amplitude of the profile related to dot 1 in the same $V_{PG2}$-range. This is because dot 2 is located next to PG2, which has a higher lever arm on dot 2 than on dot 1, leading to a relative lever arm ratio $\alpha_{PG2,dot2}/\alpha_{PG2,dot1}=2$. When sweeping $V_{PG2}$ we mainly influence dot 2.\par

We now investigate the x-position of the maximum of the profile for different $V_{PG2}$. When applying an increasingly positive voltage to PG2, we expect that dot 2 will shift to, or extend towards this gate. This becomes visible in the measurements as a lateral shift of the profiles belonging to dot 2. We observe that the maxima of the profiles of dot 1 stay at the same x-position. However, with increasing gate voltage the maxima of the profiles of dot 2 shift towards PG2 by about $20\text{ }\nano\meter/\volt$. This shift is indicated by the grey dashed line which follows the maxima of the profiles of either dot. \par

When sweeping the tip along the symmetry axis of the structure, we obtain $I(y,V_{PG2})$ as shown in Fig. \ref{fig_Grid_allGates} (b). Again we observe two sets of profiles with different amplitudes. For better visibility a profile for each dot is traced with a white line. We find that the profile belonging to dot 1 (dashed white line) has a higher amplitude than the one belonging to dot 2 (solid white line). However, their shape is the same as can be seen in the header. Furthermore, we do not observe a shift in space when sweeping the plunger gate voltage. We can conclude that the apparent y-position of the double dot is not influenced by the gate voltage applied to PG2.\par

Analogous observations are made by sweeping $V_{PG1}$ and $V_{CG}$ (not shown). The shape of the profiles of dot 1 and dot 2 remains the same for all gates evaluated as shown in Fig. \ref{fig_Grid_allGates}(c,d) for $V_{PG1}$, $V_{PG2}$ and $V_{CG}$.\par

The experimental data can be summarized as follows: First, we can observe profiles of the same shapes for both individual dots independent of the gate swept. Second, we can determine the increase in lever arm by the change in amplitude of the resonances for increasing gate voltages as well as the relative lever arms of the gates. And finally, we can quantify the apparent shift of the individual dots due to the increased voltage on a gate, which is in the order of ten to twenty nanometer per volt. \par

An obvious interpretation of this shift is, that a change in the position of the maximum of the profile allows us to measure by how much the distribution of confined electrons can be shifted in real space by an in plane gate. Although this is qualitatively correct, we have to keep in mind, that such a quantitative deduction of the shift of the electron distribution from the profile maximum might be more complex. In the presence of a scanning gate quantities such as the lever arm, charging energies and the discrete single-particle quantum levels may depend on the position of the tip \cite{Kicin_NewJPhys2005,Pioda_PRL_2004}. It is therefore hard to decide without a thorough self-consistent approximate numerical solution of the problem, which of these quantities, contributes to the apparent shift of the maxima observed in the experiment.\par

In order to make sure that the observed shift is a robust and meaningful experimental quantity we carried out z-dependent measurements.\par

To do so we recorded sets of current maps, where for each subsequent current map we scanned the tip at an increased tip-sample separation, while all voltage settings were kept the same. From each current map the $(x,y)$ coordinates of positions of both dots were extracted as follows: Each set of rings was fitted with a series of ellipses. We observe that not all ellipses have the same center point, but rather follow a linear shift. This can be attributed to the not perfectly symmetric shape of our tip. Via linear regression we determined the $(x,y)$ coordinate at which an ellipse of zero major axis length would form. This evaluation is repeated for the second set of ellipses. The parameter evaluated is the distance between the $(x_i,y_i)$ coordinate for dot 1 and dot 2, which we call the apparent dot-dot separation $s$. Determining this relative position makes the evaluation robust against shifts of the complete scan frame. The red dots in Fig. \ref{fig_Grid_allGates}(e) show the result of such a series. We see that when changing the tip-sample distance, the observed dot-dot separation remains constant as long as the tip-sample separation is below $120\text{ }\nano\meter$. For larger tip-sample separations this evaluation becomes unprecise due to the decreased number of Coulomb rings per scan frame as well as the broadening of the individual Coulomb rings in real space. This, together with the decreasing resolution of the measurements leads to a drop in the observed dot-dot separation for tip-sample distances above this value. However, all measurements presented here are carried out well below this threshold. Therefore, the observed shift is a robust experimental quantity independent of $d$ below the given limit.\par

When we carried out the same measurement and evaluation for more positive voltages applied to both plunger gates, we obtain the data set marked with blue squares. Again the dot-dot separation is independent of the tip-sample separation below a certain threshold value. For this measurement we observe an increased dot-dot separation compared to the separation extracted from the first set of current maps. The change in dot-dot separation extracted from these measurements is in the order of about $20\text{ }\nano\meter/\volt$. This is on the same order of magnitude as the shift extracted from the $I(V,position)$-maps discussed above.

\section{Conclusion}\label{sec_conclusion}
We have performed scanning gate measurements on an intentionally formed and fully controllable GaAs double quantum dot. Our experiments give spatially resolved access to the microscopic electronic properties of a double quantum dot. They allow us to locally manipulate the electron number of the individual dots. This measurement technique lets us derive the positions of both quantum dots in real space and evaluate the relative shift of their positions due to in plane gate voltages. We showed that we can also deliberately form a single dot at different positions inside the defined structure by applying different gate voltages. The position of a single dot is not altered above measurement resolution when a second dot is created next to it.\par

We would like to thank the FIRST Laboratory at the ETH Zurich for their support. We
acknowledge financial support by ETH Zurich [TH-20/05-2] and the Schweizerischer Nationalfonds.

\end{document}